\documentclass[reprint,prl,twocolumn,superscriptaddress,showpacs,nofootinbib,floatfix,preprintnumbers]{revtex4-1}
\usepackage{graphicx,bm, color, amssymb, amsmath,subfigure}

\begin{document}
\preprint{TUM-HEP-1014/15}
\title{Implications of the Diboson Excess for Neutrinoless Double Beta Decay \\ and Lepton Flavor Violation in TeV Scale Left Right Symmetric Model}

\author{Ram Lal Awasthi}
\affiliation{Indian Institute of Science Education and Research Mohali, Knowledge City, Sector 81, SAS Nagar, Manauli 140306, India}

\author{P. S. Bhupal Dev}
\affiliation{Consortium for Fundamental Physics, School of Physics and Astronomy, University of Manchester, Manchester M13 9PL, United Kingdom}
\affiliation{Physik-Department T30d, Technische Univertit\"{a}t M\"{u}nchen, 
James-Franck-Stra\ss e 1, 85748 Garching, Germany}

\author{Manimala Mitra}
\affiliation{Indian Institute of Science Education and Research Mohali, Knowledge City, Sector 81, SAS Nagar, Manauli 140306, India}
\begin{abstract}

Inspired by the recent diboson excess observed at the LHC and possible interpretation within a TeV-scale Left-Right symmetric framework, we explore its implications for low-energy experiments searching for lepton number and flavor violation. Assuming a simple Type-II seesaw mechanism for neutrino masses, we show that for the right-handed (RH) gauge boson mass and coupling values  required to explain the LHC anomalies, the RH contribution to the lepton number violating process of neutrinoless double beta decay ($0\nu\beta\beta$) is already constrained by current experiments 
for relatively low-mass (MeV-GeV) RH neutrinos. The future ton-scale $0\nu\beta\beta$ experiments could probe most of the remaining parameter space, irrespective of the neutrino mass hierarchy and uncertainties in the oscillation parameters and nuclear matrix elements. On the other hand, the RH contribution to the lepton flavor violating process of $\mu\to e\gamma$ is constrained for relatively heavier (TeV) RH neutrinos, thus providing a complementary probe of the model. Finally, a measurement of the absolute light neutrino mass scale from future precision cosmology could make this scenario completely testable. 
\end{abstract}
\maketitle
\textit {\textbf {Introduction}}-- A number of recent resonance searches with the $\sqrt s=8$ TeV LHC data have observed excess events around an invariant mass of 2 TeV. The most conspicuous one is a $3.4\sigma$  local excess in the ATLAS search~\cite{ATLAS1} for a heavy resonance decaying into a pair of Standard Model (SM) gauge bosons $VV$ (with $V=W,Z$), followed by the hadronic decay of the diboson system. The corresponding CMS search also reports a mild excess around the same mass~\cite{CMS-VV}. In addition, the CMS searches have reported a $2.2\sigma$ excess in the $WH$ channel ($H$ being the SM Higgs boson)~\cite{CMS1}, a $2.1\sigma$ excess in the dijet channel~\cite{CMS2} and a $2.8\sigma$ excess in the $eejj$ channel~\cite{CMS}, all around the same invariant mass of 2 TeV. Of course, these excesses should be  thoroughly scrutinized in light of possible subtleties in the analysis~\cite{Goncalves:2015yua} and must be confirmed with more statistics at the LHC run II before a firm conclusion on their origin could be deduced.  Nevertheless, given the lucrative possibility that it could easily be the first glimpse of new physics at the LHC, it seems worthwhile to speculate on some well-motivated beyond SM interpretations. 

One class of models that could simultaneously explain all these anomalies is the Left-Right Symmetric Model (LRSM) of weak interactions based on the gauge group $SU(2)_L\times SU(2)_R\times U(1)_{B-L}$~\cite{LR}, with the right-handed (RH) charged gauge boson mass $M_{W_R}\simeq 2$ TeV and the $SU(2)_{R}$ gauge coupling strength $g_R\simeq 0.5$ at the TeV-scale~\cite{kopp, Dobrescu:2015qna}. The main objective of this paper is  to study the implications of this scenario for various low-energy experiments searching for lepton number violation (LNV) and lepton flavor violation (LFV), which are complementary to the direct searches at the LHC. 

For concreteness, we work in the Type-II seesaw~\cite{type2} dominance, where the hitherto unknown RH neutrino mixing matrix and mass hierarchy can be directly related to the light neutrino sector. This scenario is known to give potentially sizable contributions to the low-energy LNV and LFV observables~\cite{Tello:2010, Barry:2013xxa}, apart from its novel LNV signatures at the LHC~\cite{KS, Nemevsek:2011hz}. We show that for the RH gauge boson mass and coupling values required to explain the LHC anomalies as indicated above, the LRSM parameter space for the RH neutrinos is already constrained by neutrinoless double beta decay ($0\nu\beta\beta$) experiments for relatively low (MeV-GeV) RH-neutrino masses, whereas the heavier (TeV) RH-neutrino masses are constrained by the searches for LFV processes, such as $\mu\to e\gamma$. Most of the remaining parameter space could be accessible in the next-generation $0\nu\beta\beta$ and LFV experiments at the intensity frontier. We derive a novel correlation between the $0\nu\beta\beta$ and $\mu\to e\gamma$ rates, which clearly illustrates the testability of this scenario. We further show that future  information on the absolute light neutrino mass scale from precision cosmology could make it even more predictive, irrespective of the neutrino mass hierarchy and uncertainties in the neutrino oscillation parameters or nuclear matrix elements (NMEs).    

{\textit{\textbf{Overview of the Model}}}-- Denoting
$Q\equiv (u ~~ d)^{\sf T}$ and $\psi\equiv (\nu_\ell ~~ \ell)^{\sf T}$ as the quark and lepton doublets respectively, $Q_{L}$ and $\psi_{L}$ are doublets under $SU(2)_{L}$,
 while $Q_R$ and $\psi_R$  are $SU(2)_R$ doublets. The minimal Higgs sector of the model consists of an $SU(2)_L\times SU(2)_R$ bidoublet $\Phi$ and $SU(2)_{L (R)}$-triplets $\Delta_{L (R)}$. The generic Yukawa Lagrangian of the model is given by
\begin{align}
{\cal L}_Y \  = \ & h_{q}\overline{Q}_{L}\Phi Q_{R}+\tilde{h}_{q}\overline{Q}_{L}\widetilde{\Phi} Q_{R}+
h_{l}\overline{\psi}_{L}\Phi \psi_{R}  + \tilde{h}_{l}\overline{\psi}_{L}\widetilde{\Phi}\psi_{R} \nonumber \\
&
 +f_{L} \psi^C_{L}\Delta_L \psi_{L}+f_{R}\psi^C_{R}\Delta_R \psi_{R} + {\rm H.c.},
\label{eq:yuk}
\end{align}
where $C$ stands for charge conjugation and $\widetilde{\Phi}=\tau_2\Phi^*\tau_2$ ($\tau_2$ being the second Pauli matrix). 
After electroweak symmetry breaking by the bidoublet vacuum expectation value (VEV) $\langle\Phi\rangle={\rm diag}(\kappa, \kappa')$, Eq.~(\ref{eq:yuk}) leads to the Dirac mass matrix  for neutrinos: $M_D = h_{l}\kappa + \tilde{h}_{l}\kappa'$.  The triplet VEVs $\langle\Delta^0_{L,R}\rangle=v_{L,R}$ lead to the Majorana neutrino mass terms $m_L=f_Lv_L$ and $M_R=f_Rv_R$. In the seesaw approximation, the light neutrino mass matrix 
\begin{align}
M_\nu \ \simeq \ m_L-M_DM_R^{-1}M_D^{\sf T} \; ,
\label{mass}
\end{align} 
where the first (second) term on the right-hand side is the Type-II (Type-I) seesaw contribution. An appealing case is when the Type-II contribution dominates in Eq.~\eqref{mass}, so that the smallness of the light neutrino masses is guaranteed by the smallness of $v_L \propto v^2/v_R$ (where $v=\sqrt{\kappa^2+\kappa'^2}$ is the electroweak VEV), independent of the Dirac mass matrix. Moreover, an exact $P$ (or $C$) symmetry implies $f_L=f_R$ (or $f_L=f_R^*$), so that the light and heavy neutrino mass matrices are proportional to each other, which makes it a very predictive scenario for LNV and LFV observables~\cite{Tello:2010, Barry:2013xxa}. Although in our case with $g_R<g_L$, an exact $P$ (or $C$) symmetry may not be realized down to the TeV scale, we will still work with the simple choice $f_L=f_R$, and therefore, $M_\nu=(v_L/v_R)M_R$. Setting $M_{W_R}\simeq 2$ TeV and $g_R\simeq 0.5$, as required to explain the LHC diboson and dijet anomalies~\cite{kopp, Dobrescu:2015qna}, also fixes the $SU(2)_R$ breaking scale $v_R\simeq 6$ TeV. Note that for purely Majorana RH neutrinos as in the minimal LRSM, it is difficult to fit the CMS $eejj$ excess~\cite{CMS}, which requires a suppression of the same-sign dielectron signal and the absence of an $ejj$ peak. For possible alternatives, see e.g.~\cite{Dobrescu:2015qna, Gluza:2015goa, Saavedra:2015rna, Krauss:2015nba}. 

{\textit{\textbf{Neutrinoless Double Beta Decay}}}-- In the LRSM, there are several new contributions to the $0\nu\beta\beta$ amplitude~\cite{review}, apart from the canonical light neutrino exchange diagram.  However, in the Type-II seesaw dominance, all the mixed LH-RH contributions are negligible, whereas the scalar triplet contribution can also be neglected for $M_R/M_\Delta\lesssim 0.1$~\cite{Tello:2010}, which is assumed here.  Thus, we are only left with the purely LH- and RH-contributions to the $0\nu\beta\beta$ half-life: 
\begin{eqnarray}
\frac{1}{T_{1/2}^{0\nu}} \ = \ G_{0\nu}g_A^4|{\cal M}_\nu|^2\left|\frac{m_{ee}^{\nu}+m_{ee}^N}{m_e}\right|^2 , 
\label{halft}
\end{eqnarray}
where $G_{0\nu}$ is the phase space factor, $g_A$ is the nucleon  axial-vector coupling constant, $m_e$ is the electron mass, ${\cal M}_{\nu, N}$ are the NMEs and $m_{ee}^{\nu, N}$ are the effective neutrino masses corresponding to light and heavy neutrino exchange, respectively. For the light neutrino exchange, $m_{ee}^\nu=\sum_i U^2_{ei}m_i$, $U$ being the PMNS mixing matrix which diagonalizes the light neutrino mass matrix $M_\nu$ with eigenvalues $m_i$. Using the standard parametrization for $U$ in terms of three mixing angles $\theta_{ij}$, one Dirac $CP$ phase $\delta$ and two Majorana $CP$ phases $\alpha_{2,3}$, we get 
\begin{eqnarray}
m^\nu_{ee} \ = \ m_1c_{12}^2c_{13}^2+m_2s_{12}^2c_{13}^2e^{2i\alpha_2}+m_3s_{13}^2e^{2i\alpha_3} ,
\label{mnuee}
\end{eqnarray}  
where $c_{ij}\equiv \cos \theta_{ij}$ and $s_{ij}\equiv \sin \theta_{ij}$ (for $i,j=1,2,3$).

For the heavy neutrino exchange due to RH current, 
\begin{eqnarray}
m_{ee}^N \ = \ |p^2|\left(\frac{g_R}{g_L}\right)^4 \left(\frac{M_{W_L}}{M_{W_R}}\right)^4\sum_j\frac{V_{ej}^2 M_j}{|p^2|+M_j^2}  \, ,
\label{mNee}
\end{eqnarray} 
where $|p^2| = m_em_p{{\cal M}_N}/{{\cal M}_\nu}\sim (100~{\rm MeV})^2$ denotes the typical momentum exchange, $m_p$ is the proton mass, and $V_{ej}$ are the elements of the first row of the unitary matrix diagonalizing the RH neutrino mass matrix $M_R$ with eigenvalues $M_j$. Since $m_i\propto M_i$ in the Type-II dominance, for normal hierarchy (NH), $M_3$ will be the largest (henceforth denoted as $M_>$), and we can express the other two RH neutrino masses as $M_1=(m_1/m_3)M_>$ and $M_2=(m_2/m_3)M_>$. 
Similarly, for inverted hierarchy (IH), $M_2$ will be the largest and we can write $M_1=(m_1/m_2)M_>$ and $M_3=(m_3/m_2)M_>$. 

\begin{figure*}[t!]
\includegraphics[width=16cm]{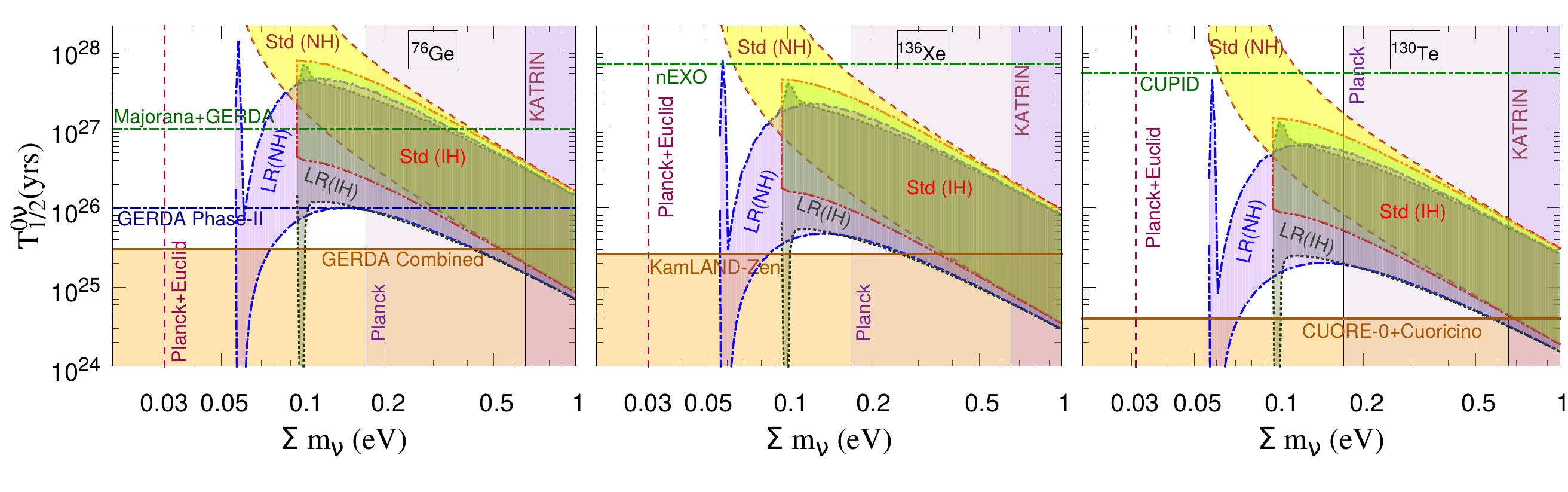}
\caption{The $0\nu\beta\beta$ half-life predictions as a function of the sum of light neutrino masses for different isotopes in our LRSM scenario with the largest RH neutrino mass $M_>=1$ TeV for both NH and IH. For comparison, the canonical light neutrino contributions (Std) are also shown. The current 90\% CL experimental limits~\cite{Agostini:2013mzu, Gando:2012zm, Alfonso:2015wka} and future sensitivities~\cite{Majorovits:2015vka, Abgrall:2013rze, Albert:2014afa, Wang:2015raa} are shown for each isotope (horizontal lines). The vertical lines (from right to left) show the future 90\% CL limit from KATRIN~\cite{Mertens:2015ila}, current 95\% CL limit from Planck~\cite{Ade:2015xua} and future 95\% CL limit from Planck+Euclid~\cite{Cerbolini:2013uya}.}
\label{fig2}
\end{figure*}

For the numerical analysis, we consider three relevant isotopes, namely, $^{76} \rm{Ge}$, $^{136}\rm{Xe}$ and $^{130}\rm{Te}$ (see e.g.~\cite{Pandola:2014naa} for the status of experiments with other isotopes), and compare the LRSM predictions for the half-life with the corresponding current experimental limits and future sensitivities. For the phase space factors and NMEs, we use the SRQRPA calculations with $g_A=1.25$ from~\cite{Meroni:2012qf}. Our results are shown in Fig.~\ref{fig2} as a function of the sum of light neutrino masses for an illustrative value of $M_>=1$ TeV. In each case, we show the LRSM predictions (LR) for both NH and IH, and the corresponding light neutrino contributions alone (Std) for comparison. The kink in the LR contribution is due to a cancellation in the $0\nu\beta\beta$ amplitude when the lightest neutrino mass becomes very small: $m_{\rm{lightest}} \lesssim 1~\mu$eV.
Here we have used the $3\sigma$ allowed ranges of the neutrino mass and mixing parameters from a recent global fit~\cite{Garcia:2014bfa} and have varied the Majorana phases $\alpha_{2,3}\in [0,\pi]$. The lower horizontal lines show the current 90\% CL limits on the half-life from GERDA combined with Heidelberg-Moscow and IGEX for $^{76}$Ge~\cite{Agostini:2013mzu}, KamLAND-Zen for $^{136}$Xe~\cite{Gando:2012zm} (see also EXO-200~\cite{Albert:2014awa}), and Cuore-0 combined with Cuoricino for $^{130}\rm{Te}$~\cite{Alfonso:2015wka}. The projected limits from GERDA phase-II~\cite{Majorovits:2015vka} as well as the planned ton-scale experiments such as MAJORANA+GERDA~\cite{Abgrall:2013rze}, nEXO~\cite{Albert:2014afa}, and CUPID~\cite{Wang:2015raa} are shown by the upper horizontal lines. The rightmost vertical line in each plot represents the future 90\% CL sensitivity of KATRIN~\cite{Mertens:2015ila} for the absolute neutrino mass scale, whereas the other two vertical lines show the best 95\% CL limit on the sum of light neutrino masses from Planck data~\cite{Ade:2015xua} and a projected 95\% CL limit from Planck+Euclid~\cite{Cerbolini:2013uya}.

From Fig.~\ref{fig2}, it is evident that the future sensitivity reach of the ton-scale $0\nu\beta\beta$ experiments can completely probe this LRSM benchmark point, 
  irrespective of the neutrino mass hierarchy and uncertainties due to oscillation parameters and NMEs, except for the cancellation region in the NH case. This can be further constrained by the cosmological limit on the sum of light neutrino masses~\cite{Dell'Oro:2015tia, Ge:2015yqa}. In particular, the precision cosmology with the future Euclid project could provide an indirect measure of the absolute neutrino mass, thus possibly ruling out the cancellation region and enabling more definitive model predictions for $0\nu\beta\beta$.

\begin{figure*}[t!]
\includegraphics[width=7cm]{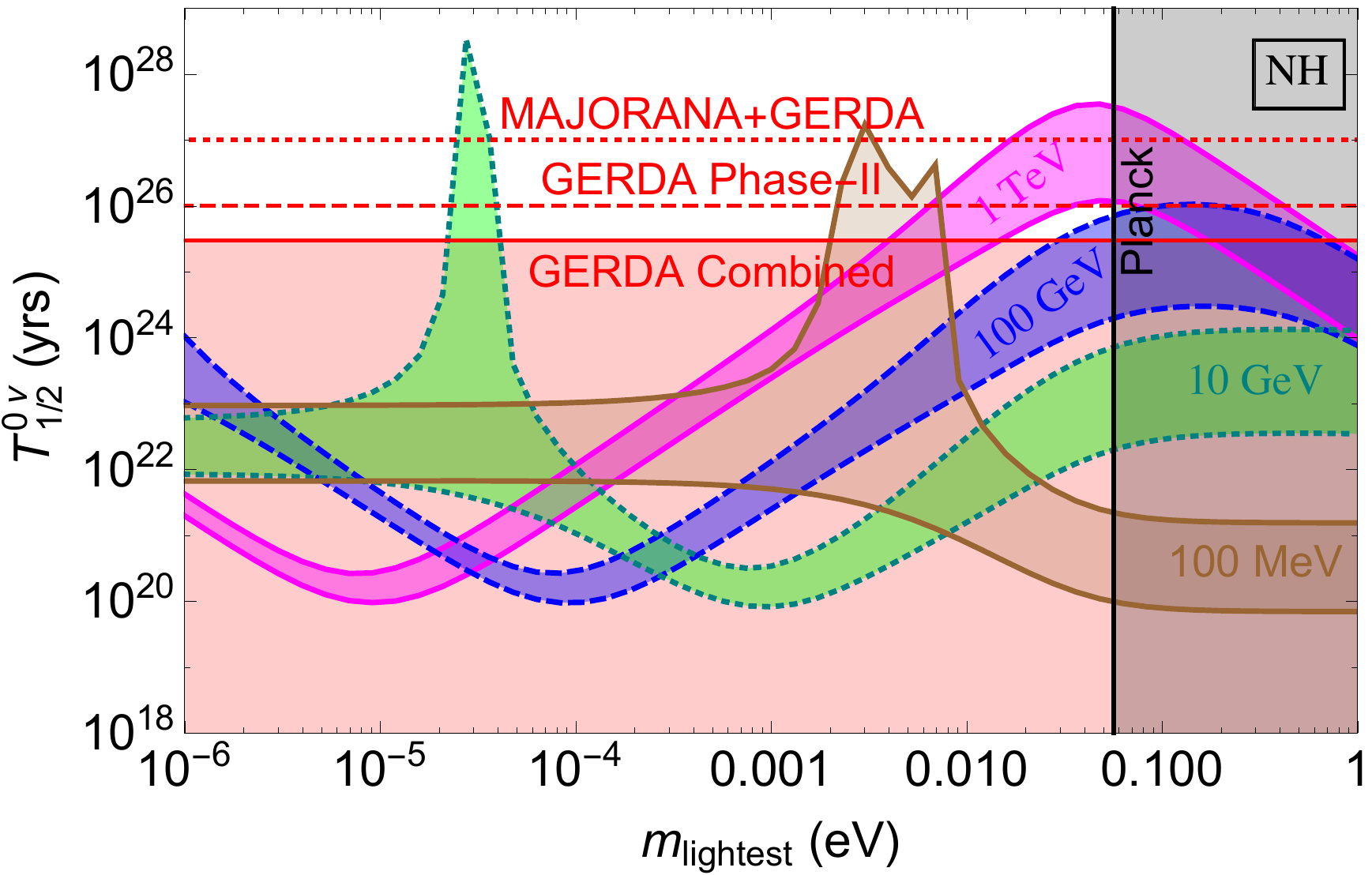}
\includegraphics[width=7cm]{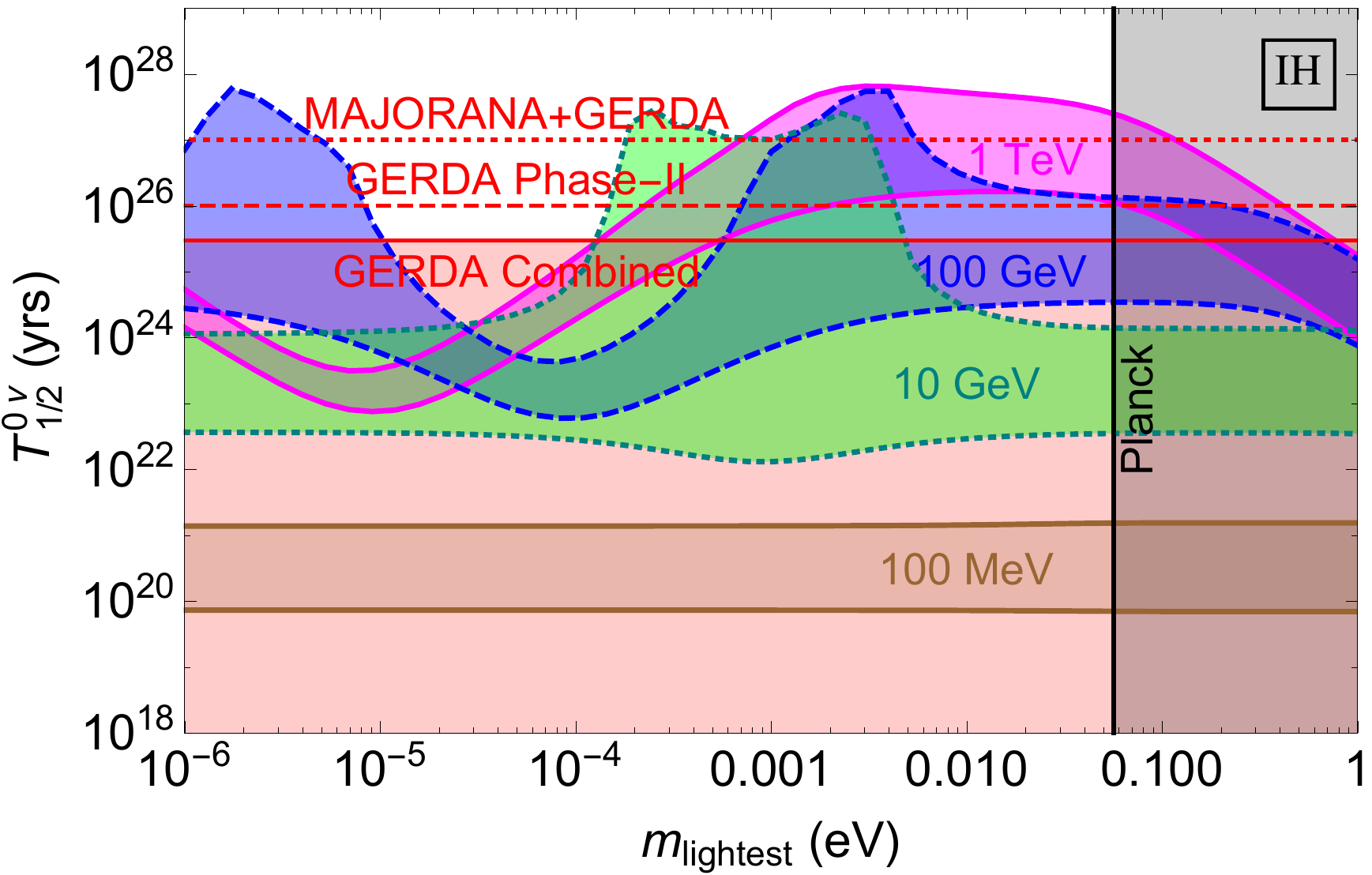}
\caption{The $0\nu\beta\beta$ half-life of $^{76}$Ge as a function of the lightest neutrino mass for various values of $M_>$ in our LRSM scenario.}
\label{fig3}
\end{figure*}

The variation of the $0\nu\beta\beta$ half-life predictions in our LRSM scenario with respect to the heavy neutrino mass parameter $M_>$ is examined in Fig.~\ref{fig3} for both NH and IH. For illustration, we have only shown the results for $^{76}$Ge isotope and have varied the lightest neutrino mass between $10^{-6}$--1 eV. Note that in the NH case, there always exists a cancellation region for the LR contribution by itself, which shifts to smaller $m_{\rm lightest}$ values as we increase the RH neutrino mass scale $M_>$. This is reminiscent of the NH cancellation region for the canonical light neutrino contribution alone, which occurs between $m_{\rm lightest}=1$--10 meV. On the other hand, for the IH case, the cancellation in the LR contribution occurs only when at least one of the RH neutrino masses is above $|p^2|\sim (100~{\rm MeV})^2$, in contrast with the light neutrino case, where there is no cancellation for IH. We do not show the results for $m_{\rm lightest}<1~\mu$eV, because in the Type-II seesaw dominance, this will imply the lightest RH neutrino mass close to eV (for some of the benchmark values of $M_>$ shown in Fig.~\ref{fig2}), which is already disfavored by cosmology~\cite{Ade:2015xua}. From Fig.~\ref{fig3}, we find that the $M_>=100$ MeV IH case is already ruled out by GERDA phase-I~\cite{Agostini:2013mzu}, and in the NH case, only the cancellation region survives. For higher $M_>$ values with both NH and IH, part of the parameter space is already ruled out by GERDA, and the remaining can be probed in future~\cite{Majorovits:2015vka, Abgrall:2013rze}, except the cancellation regions. 
Further information on the absolute light neutrino mass scale could in principle eliminate these cancellation regions. 

{\textit{\textbf{Lepton Flavor Violation}}}-- There exist several Feynman diagrams contributing to the LFV observables such as $\ell \to \ell' \gamma$, $\ell \to 3\ell'$ and $\mu-e$ conversion.   Focusing on the most promising LFV process $\mu \to e\gamma$, the purely RH contribution to the branching ratio is given by  
\begin{align}
\textrm{Br}({\mu \to e \gamma}) \simeq \frac{3\alpha_{\rm em}}{2\pi} \left(\frac{g_R}{g_L}\right)^4\left(\frac{M_{W_L}}{M_{W_R}} \right)^4 \left|\sum_i V_{\mu i}V^*_{ei}G_\gamma(x_i)\right|^2,
\label{mu2eg}
\end{align}
where $\alpha_{\rm em}=e^2/4\pi$ is the electromagnetic coupling constant, $x_i\equiv (M_i/M_{W_R})^2$ and the loop-function 
\begin{align}
 G_{\gamma}(x) \ = \ -\frac{2x^3+5x^2-x}{4(1-x)^3} - \frac{3x^3}{2(1-x)^4}\ln{x} ,
\end{align}
which approaches the constant value of 1/2 in the limit $x\gg 1$. Note that the SM contribution to $\textrm{Br}({\mu \to e \gamma})$ due to light neutrino exchange is extremely small $\lesssim 10^{-55}$~\cite{Bilenky:1977du}, and furthermore, the other LH, mixed and scalar contributions can be neglected in our LRSM scenario under the assumptions stated before. 
\begin{figure}[t!]
\includegraphics[width=7cm]{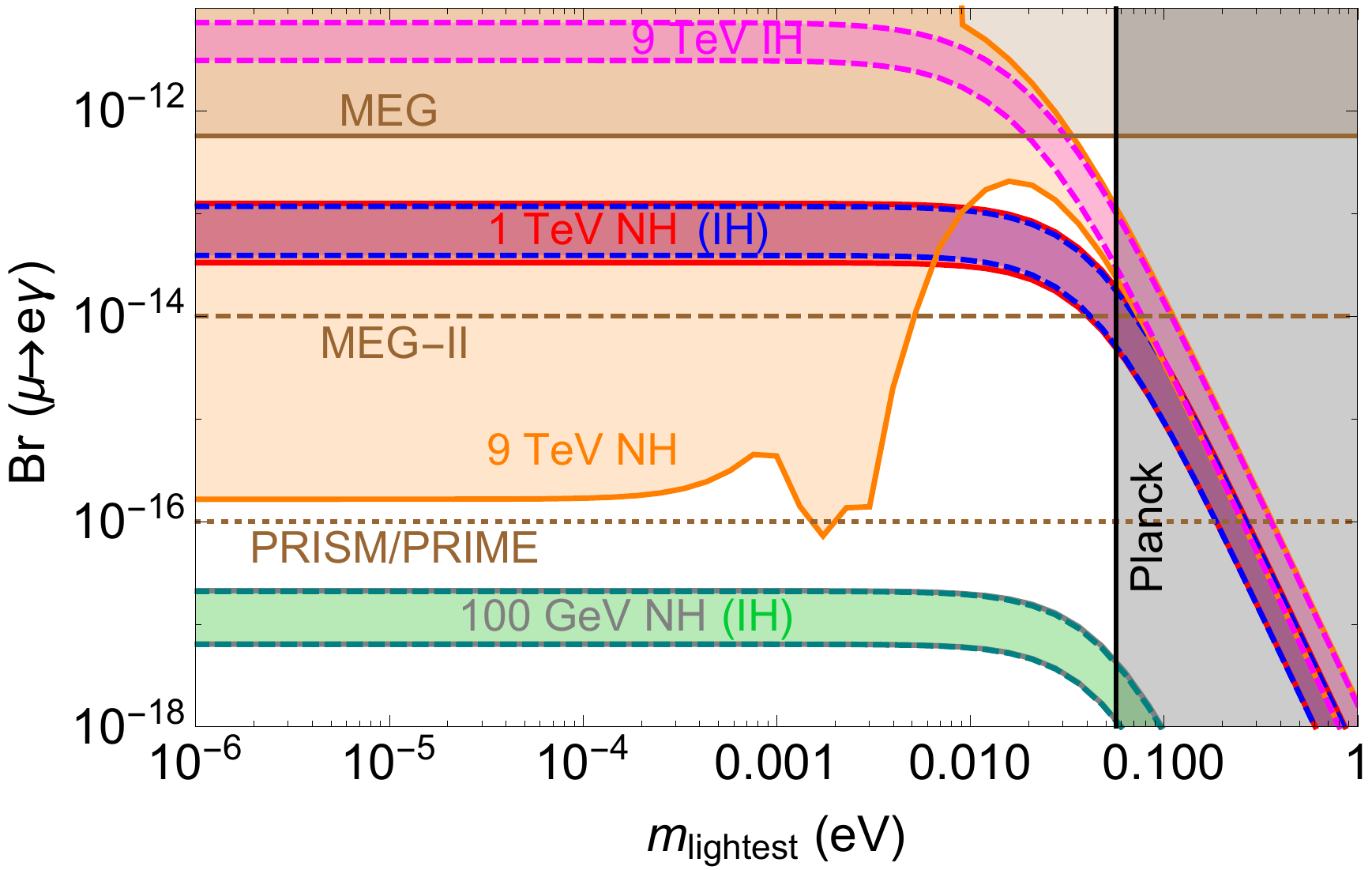}
\caption{Branching ratio of $\mu \to e\gamma$ as a function of the lightest neutrino mass in our LRSM with different values of $M_>$.}
\label{fig4}
\end{figure}

Our predictions for Br$(\mu\to e\gamma)$ from Eq.~\eqref{mu2eg} are shown in Fig.~\ref{fig4} for both NH and IH with three benchmark values of $M_>$, where the band in each case is due to the $3\sigma$ uncertainties in the neutrino oscillation parameters. The horizontal shaded region is ruled out at 90\% CL by the MEG experiment~\cite{Adam:2013mnn}, while the horizontal dashed and dotted lines show the MEG-II~\cite{Baldini:2013ke} and PRISM/PRIME~\cite{Kuno:2005mm} sensitivities, respectively. It is evident that the $\mu\to e\gamma$ searches could be more effective in probing the relatively heavier $M_>$ values, as compared to the $0\nu\beta\beta$ searches. Also note that the Planck upper limit on the lightest neutrino mass effectively puts a lower limit on the $\mu\to e\gamma$ branching ratio in the quasi-degenerate (QD) region; for instance, for $M_>=1$ TeV, this lower limit is $(0.5-1.9)~ [(0.5-1.7)] \times 10^{-14}$ for NH [IH], as shown in Fig.~\ref{fig4}. However, for NH with $M_>/M_{W_R}\gtrsim 1$, there is a destructive interference between the two heaviest neutrino contributions for certain values of the Dirac $CP$ phase, which leads to a cancellation region, unless the third RH neutrino contribution is sizable, as demonstrated in Fig.~\ref{fig4} for $M_>=9$ TeV. 
On the other hand, smaller $M_>$ values lead to a suppression in the $\mu\to e\gamma$ rate, pushing it well below the  future sensitivity even for $M_>=100$ GeV, which is however accessible to $0\nu\beta\beta$ experiments. Thus, a combination of the low-energy probes of LNV and LFV is crucial to probe effectively the entire LRSM parameter space in our case. 
This is complementary to the direct searches at the LHC~\cite{CMS, Aad:2015xaa}, which can probe RH neutrino masses from about 100 GeV upto $< M_{W_R}$~\cite{Ferrari:2000sp} using the same-sign dilepton plus dijet channel~\cite{KS}. Similarly, the GeV-scale RH neutrinos can also be searched for in the proposed SHiP experiment~\cite{Felisola:2015bha}.   

\begin{figure}[t!]
\includegraphics[width=7cm]{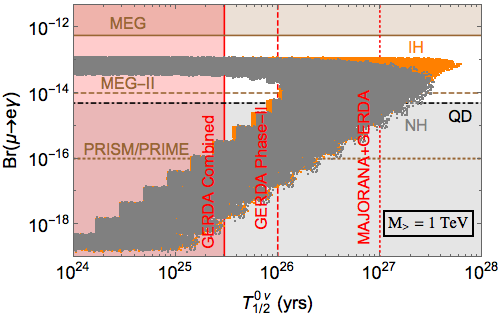}
\caption{Correlation between $0\nu\beta\beta$ and $\mu\to e\gamma$ in our LRSM scenario for  $M_>=1$ TeV. }
\label{fig5}
\end{figure}

{\textit{\textbf{Correlation between LNV and LFV}}}-- The synergistic aspects of the low-energy LNV and LFV searches is further illustrated in Fig.~\ref{fig5} via a correlation between the $0\nu\beta\beta$ half-life of $^{76}$Ge and $\mu\to e\gamma$ branching ratio in our LRSM setup. This plot is obtained for a typical value of $M_>=1$ TeV, while the scattered points are due to the $3\sigma$ variation  of the oscillation parameters and the NME uncertainties. The upper horizontal shaded area is ruled out by the MEG experiment~\cite{Adam:2013mnn}, whereas the vertical shaded area is excluded by GERDA phase-I~\cite{Agostini:2013mzu}. The lower horizontal shaded area corresponds to QD light neutrino masses, which is disfavored by Planck~\cite{Ade:2015xua}, as also shown in Fig.~\ref{fig4} by the vertical shaded area. All the remaining region can in principle be probed by a combination of the future $0\nu\beta\beta$ and LFV experiments, as shown by the dashed/dotted lines in Fig.~\ref{fig5}.              

{\textit{\textbf{Conclusion}}}-- We have explored the low-energy implications of the recent diboson excess observed at the LHC in the context of a TeV-scale Left-Right Symmetric model with Type-II seesaw dominance. In particular, we analyze the predictions for $0\nu\beta\beta$ and $\mu\to e\gamma$, and show that a combination of these experiments at the intensity frontier can effectively probe most of the hitherto unknown RH neutrino parameter space of this LRSM scenario. We find that the RH neutrinos with relatively low mass (MeV-GeV) are already ruled out from the existing bound on $0\nu \beta \beta$, apart from some cancellation regions, whereas the future ton-scale $0\nu \beta \beta$ experiments could probe most of the remaining parameter space of this model. On the other hand, the TeV-scale RH neutrinos in this scenario are constrained by the MEG limits on $\mu \to e \gamma$ decay rate. The synergistic aspects of the future LNV and LFV experiments at the intensity frontier is demonstrated by a novel correlation between the $0\nu\beta\beta$ half-life and the $\mu\to e\gamma$ branching ratio. Finally, a measurement of the absolute neutrino mass scale from future precision cosmology could render the model predictions for LNV and LFV more definitive. 

\acknowledgements
{\textit{\textbf{Acknowledgments}}}-- P.S.B.D. acknowledges the local hospitality and partial support from the European Centre for Theoretical Studies in Nuclear Physics and Related Areas (ECT*) during the completion of this work. The work of P.S.B.D. is supported in part by a TUM University Foundation Fellowship and the DFG cluster of excellence ``Origin and Structure of the Universe". M.M would like to acknowledge the generous support provided by IISER Mohali  and unacknowledge the DST-INSPIRE Faculty Scheme for not providing any support.

\end{document}